\documentclass{PoS}
\usepackage{siunitx}
\usepackage{lineno}

\title{From gated to continuous readout $\mathbf{-}$ the GEM upgrade of the ALICE TPC}

\ShortTitle{From gated to continuous readout $-$ the GEM upgrade of the ALICE TPC}

\author{\speaker{Andreas Mathis} on behalf of the ALICE collaboration\\
        Physik Department E62, Technische Universit\"at M\"unchen, Garching.\\
        Excellence Cluster 'Origin and Structure of the Universe', Garching.\\
        E-mail: \email{andreas.mathis@ph.tum.de}}
\abstract{
The ALICE Collaboration is planning a major upgrade of its central barrel detectors to be able to cope with the increased LHC luminosity beyond 2020. In order to record at an increased interaction rate of up to 50\,kHz in Pb$-$Pb collisions, the TPC will be operated in an ungated mode with continuous readout. This demands for a replacement of the currently employed gated Multi-Wire Proportional Chambers by GEM-based (Gas Electron Multiplier) readout chambers, while retaining the performance in particular in terms of particle identification capabilities via the measurement of the specific energy loss.

Prior to the beginning of the full mass production of the readout chambers for the upgrade, a so-called pre-production was launched in order to characterize and verify the performance of the first fully assembled readout chambers of the final design.
This phase was concluded in March 2017 with the formal acceptance of the production readiness and hence the beginning of the mass production.
}

\FullConference{5th International Conference on Micro-Pattern Gas Detectors (MPGD2017)\\
		22-26 May, 2017\\
		Philadelphia, USA}

\begin{document}

\section{Introduction}
A significant increase of the LHC instantaneous luminosity of up to $6\times10^{27}$\,cm$^{2}$s$^{-1}$ will enable the second generation of LHC heavy-ion studies beyond 2020, after the so called Long Shutdown 2 (LS2). In order to fully exploit this increased luminosity, a continuous readout of the ALICE detector at an interaction rate of 50\,kHz in Pb$-$Pb is foreseen.
The ALICE Time Projection Chamber \cite{ALICETPC}  is the main device for charged-particle tracking, momentum measurement and particle identification (PID) in the central barrel of the ALICE detector. It consists of a hollow cylindrical barrel with a central cathode and two readout planes on each end cap which are azimuthally segmented into 18 sectors, each covering 20$^{\circ}$. A further division of each sector into Inner (IROC) and Outer Readout Chamber (OROC) is motivated by the different requirements for the readout chambers as a function of the radius due to the radial dependence of the track density. With an overall volume of about 90\,m$^{3}$ it is the largest detector of its kind to date. The presently employed gated Multi-Wire Proportional Chambers, in combination with the readout electronics, allow for a readout rate as high as $\sim$3\,kHz. In Run 3, however, the interaction rate will be increased to 50\,kHz implying an average event pileup of 5. A triggered operation of the TPC with a gating grid will then no longer be possible, as it would cause inacceptable losses of data. 

The solution for the TPC upgrade \cite{TPC1,TPC2} consists of a stack of four large-size Gas Electron Multiplier (GEM) \cite{GEM} foils as amplification stage. This arrangement, under a specific high-voltage configuration, has been proven to fully meet the design specifications in terms of ion backflow (IBF), energy resolution and stable operation under LHC conditions. In order to further guarantee operational stability for the readout chambers during construction, commissioning and final operation in the TPC, a sophisticated quality assurance scheme has been established to thoroughly monitor the quality of the GEM foils throughout the whole production process.
Prior to the beginning of the mass production, two readout chambers of the final design were assembled and qualified.

\section{Final GEM design}
The GEM foils for the ALICE TPC upgrade are produced by the CERN PCB workshop using the single-mask technique \cite{single-mask} according to the final design described in the following. Their trapezoidal shape is given by the size and shape of the presently installed readout chambers. Due to technical limitations on the size of the processed raw material, the outer readout chambers are equipped with three independent GEM stacks, as summarized in Tab.~\ref{tab:GEMsize}. 
According to the baseline configuration \cite{TPC2} the GEM stacks contain both Standard (S, pitch \SI{140}{\micro\meter}) and Large Pitch foils (LP, pitch \SI{280}{\micro\meter}) in the order S-LP-LP-S.

The top side of each foil is segmented such that the individual so-called HV segments have an area of about $100$\,cm$^{2}$, in order to limit the current flowing in case of potential electrical discharges.
The gap between the copper layers of adjacent segments is \SI{200}{\micro\meter}, with an additional \SI{100}{\micro\meter} space between the segment boundaries and the GEM holes to accommodate possible misaligments during foil production. Each segment is connected via a  \SI{5}{\mega\ohm} SMD\footnote{Surface-mount device} 206 loading resistor to the common HV distribution, which is a \SI{1}{\milli\meter} wide copper path running along three sides of the foil.
These SMD pads are located parallel to the segment boundary with dimensions defined according to the specifications of the producer (CERN PCB workshop): a pad size of $1.6 \times 2$\,mm$^{2}$ with a 1.2\,mm gap between the individual pads.
Directly after production, the GEM foils are equipped with the loading resistors and the solder is cleaned in order to remove corrosive content of the solder.

\begin{figure}[t]
\centering
\includegraphics[width = 0.95\textwidth]{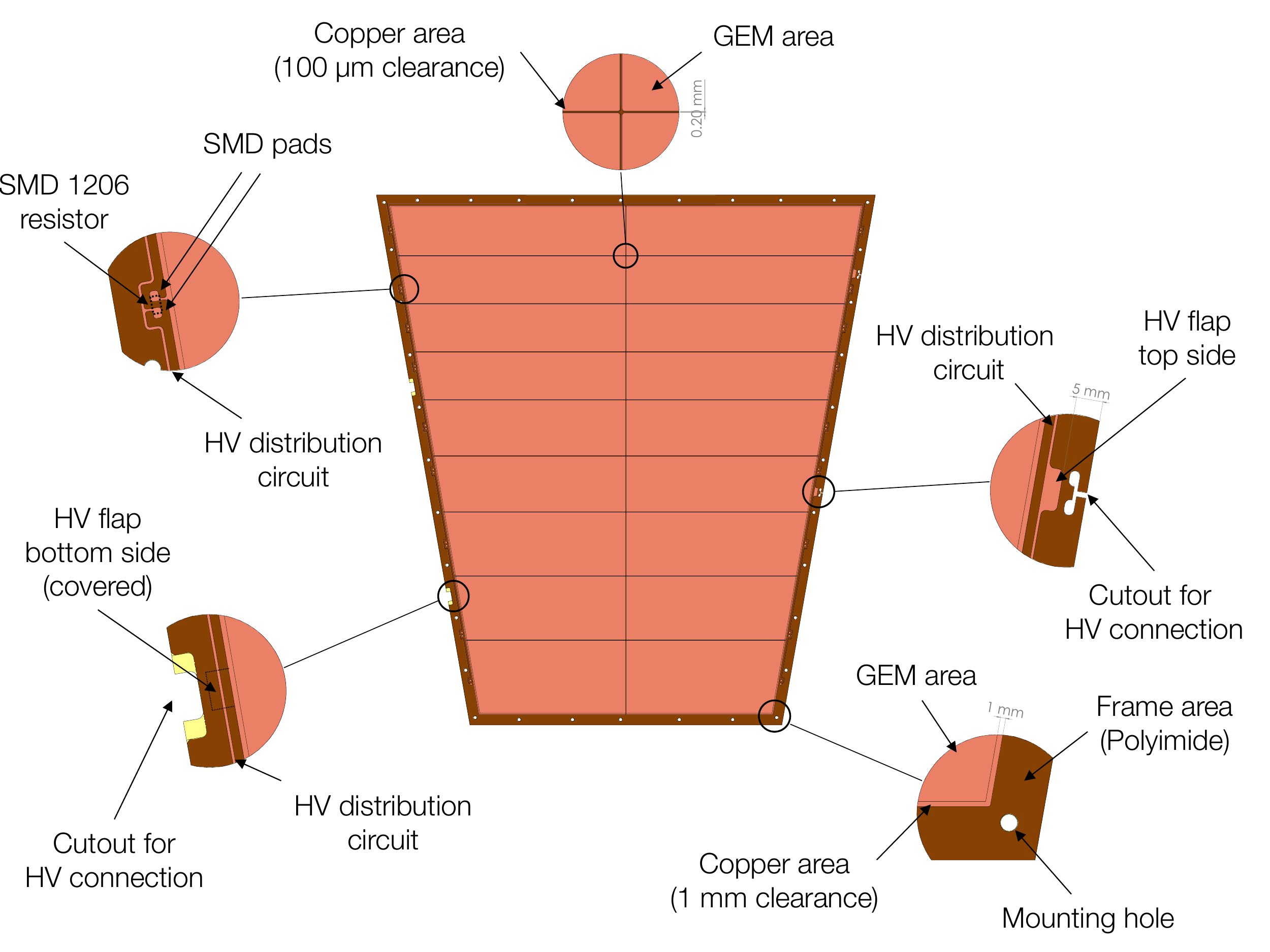}
\caption{Overview of the most important elements of the final GEM design. For details see text.}
\label{fig:GEMdesign}
\end{figure}

Particularly important is to maximize the distance of any electrically live element from the readout chamber boundary, because in the worst case of a HV trip in a neighboring sector, a large potential difference of up to 4\,kV may occur between the two sectors. The maximal possible distance between the HV distribution to the chamber boundary is \SI{5}{\milli\meter}, which, taking into account the distance between two adjacent readout chambers, accounts to a total minimal distance of electrically live elements of \SI{13}{\milli\meter}. Tests in the laboratory with a specifically designed mock-up proved this distance to be safe.
In order to reduce the electric field at all boundaries, no radii smaller than \SI{0.5}{\milli\meter} are allowed.
Figure~\ref{fig:GEMdesign} displays the most important features of the final GEM design for the ALICE TPC upgrade.

\begin{table}[t]
\begin{center}
\caption{Size and important characteristics of the GEM foils. The width of the chambers is indicated by the long (short) side of the trapezoid.}
\label{tab:GEMsize}
\begin{tabular}{ c |c | c | c }
GEM type & Size &  \# HV  & avg. segment size  \\
 & H (cm) $\times$ W (cm) & segments & (cm$^{2}$) \\
 \hline
 IROC & $49.7 \times 46.7 \,(29.2)$ & 18 & 93 \\
 OROC 1 & $36.2 \times 59.5 \,(46.8)$ & 20 & 87 \\
 OROC 2 & $38.0 \times 73.0 \,(59.6)$ & 22 & 105 \\
 OROC 3 & $39.8 \times 87.0 \,(73.0)$ & 24 & 122\\
\end{tabular}
\end{center}
\end{table}

\section{Pre-production}
The successfully completed prototyping phase of the project, summarised in \cite{RD1, RD2, RD3, RD4} was followed by the so-called pre-production of the GEM readout chambers. The goal of this phase was to assemble and characterize one readout chamber of each type, i.e.\ one IROC and one OROC, in order to verify details of the final design, finalize the quality assurance (QA) and production procedures and protocols.
A full characterization of these readout chambers was conducted to prove conformance with the challenging requirements of the upgrade.

The work flow of the upgrade is as follows. After production and a first QA at CERN, the foils are distributed to the so-called Advanced QA centers (Helsinki Institute of Physics, Wigner Research Center for Physics Budapest), followed by the framing centers, in which the GEMs are glued onto their supporting frames (Rheinische Friedrich-Wilhelms-Universit\"at Bonn, GSI Helmholtzzentrum f\"ur Schwerionenforschung, Technical University of Munich, Yale University). The assembled GEMs are then sent to the assembly centers (National Institute for Physics and Nuclear Engineering Bucharest, GSI, Yale), where the GEMs are mounted onto the readout chamber body. The final detector is then qualified and sent to CERN for storage and final integration in the year 2019.

\subsection{Quality assurance}
Details of the QA procedures and protocols are discussed in \cite{QA2, QA}. In order to reject malfunctioning GEM foils at the earliest possible stage, all GEM foils are, prior to each production step, subject to a coarse optical inspection and to a measurement of the leakage current. The latter is conducted using a multi-channel HV system with a maximally acceptable leakage current of \SI{0.5}{\nano\ampere} per HV segment.
After the production and a first characterization at CERN, the foils are shipped to the Advanced QA centres, where by means of high-definition optical scanning the foils are characterized. This concerns in particular a detailed mapping of the full foil regarding the hole size distribution and defect classification \cite{QA}.

\subsection{Assembly overview}
The assembly steps of the readout chambers, after minor modifications, follow the procedures used for the production of the first IROC and OROC prototypes as discussed in \cite{RD2, RD4}.
%The ROC assembly consists of two mains steps, the GEM framing and the ROC assembly.

\subsubsection{GEM framing}
The first step of the GEM framing consists of the assembly of the frames from four ledges made of \SI{2}{\milli\meter} thick Vetronite EGS 103 (reinforced fiber glass). Additionally, a \SI{1.5}{\milli\meter} wide spacer cross is mounted serving to counteract the electrostatic attraction of adjacent GEM foils.
Grooves milled into the bottom side of the frame accommodate the loading resistors of the GEM foil mounted below. Therefore, framed foils can be placed flat on top of each other and the distance between two foils in the stack is solely given by the thickness of the frame.

Before being glued onto the pre-assembled frame, the foils are pneumatically stretched with a tension of 10\,N/cm using a modified commercial tool. The frame is covered with the epoxy glue ARALDITE 2011\textsuperscript{TM} and thoroughly inserted into the gluing jig. The foil is then positioned on the frame and aligned using alignment pins and the jig itself. The gluing jig is closed with a heavy cover plate milled such that the active area of the foil is not touched and burdened with steel bricks to press the foil onto the frame. The glue is cured for 24 hours in dry atmosphere (below \SI{30}{\percent} relative humidity). After that, the foil is removed from the gluing jig and sent to the assembly center for further processing.

\subsubsection{ROC assembly}
After a final measurement of the leakage current, the raw material surrounding the active area of the GEM foils is removed.
Then, the foils are subsequently mounted on the readout chamber bodies, which consist of the pad plane, a multi-layer Printed Circuit Board (PCB), an additional \SI{5}{\milli\meter} Vetronite EGS 103 insulation plate ('strong back') and an aluminum frame (alubody).
The stack is fixed to the alubody using reinforced nylon screws. 
In a last step the PEEK-coated wires are soldered to the HV flaps on each foil in order to provide high voltage.

\subsection{Full ROC characterization}
%Before storage and integration of the newly built readout chambers, a set of thorough characterization tests is conducted.

\subsubsection{Gas tightness}
The fully assembled readout chamber is mounted to a test box and flushed with the nominal detector gas (Ne-CO$_{2}$-N$_{2}$ (90-10-5)), while monitoring the oxygen  and humidity contamination of the exhausted gas. Once the oxygen level has reached its asymptotic value, the leak rate is determined and should be below 0.5\,ml/h (e.g.\ $<$\,5\,ppm oxygen at a flow of 20\,l/h) at close to atmospheric pressure.

\subsubsection{Gain curve}
\begin{figure}[t]
\centering
\includegraphics[width=0.65\textwidth]{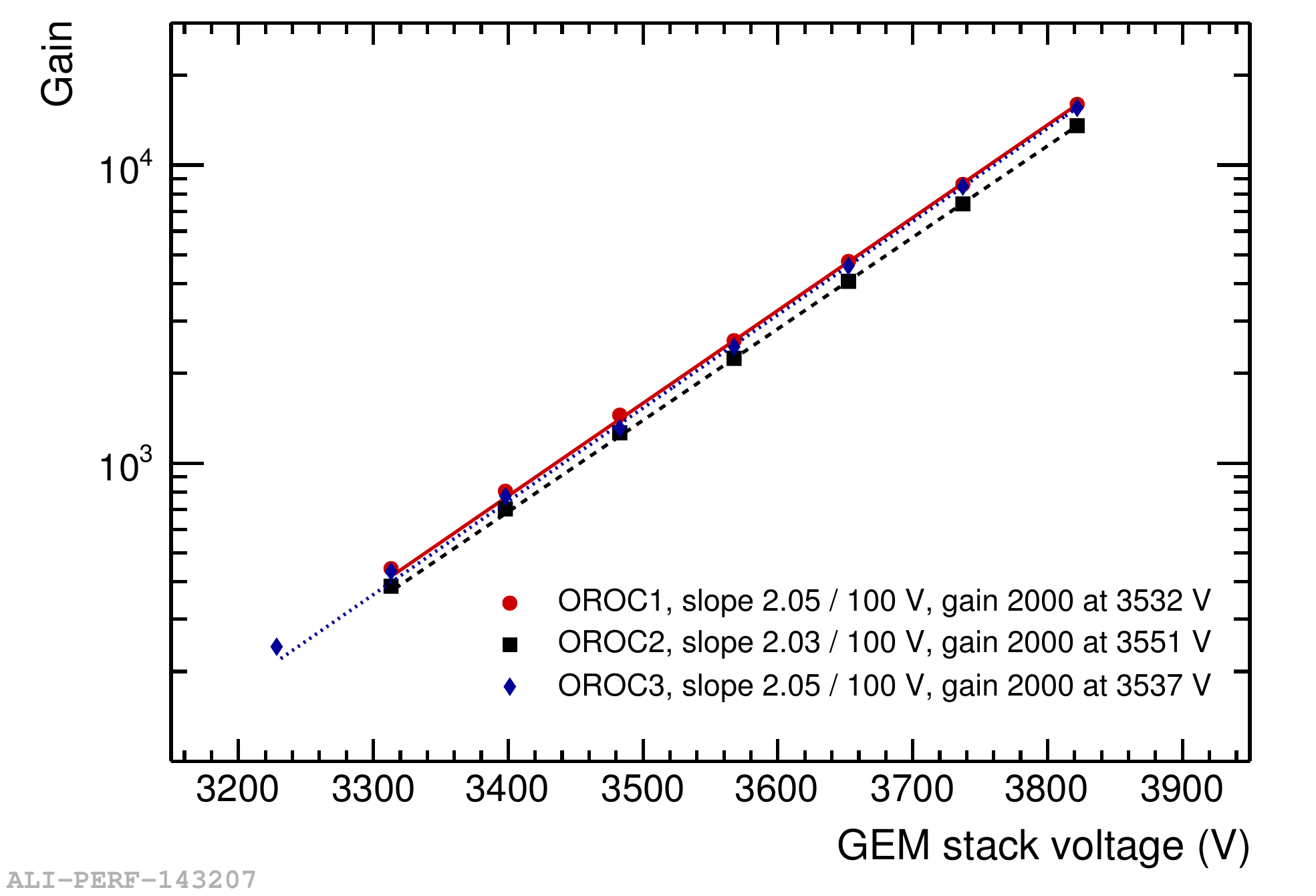}
\caption{Gain curve measured for the three OROC GEM stacks. }
\label{fig:OROCgain}
\end{figure}

In order to determine the gain curve of the fully assembled readout chamber, the latter is mounted to a test box including a drift cathode. High voltage is applied to the GEM foils via a voltage divider, subsequently degrading the potential for each electrode. The HV configuration applied to the GEM stack in the course of this test is displayed in Tab.~\ref{tab:GEMHV}.
The gain is then determined by measuring the rate of a collimated $^{55}$Fe source and the corresponding current induced on the readout anode. For a given chamber, the overall voltage applied to the voltage divider is varied such that the gain of the chamber is scanned between $\sim$1000 and $\sim$10000. The results of this measurement for the pre-production OROC is displayed in Fig.~\ref{fig:OROCgain}.
Additionally, a spectrum of the $^{55}$Fe source is recorded in every stack at nominal gain in order to quantify the corresponding energy resolution.

\begin{table}[t]
\begin{center}
\caption{HV setting used for the gain characterization.}
\label{tab:GEMHV}
\begin{tabular}{ c |c | c | c |c |c |c |c  }
$\Delta U_{\mathrm{GEM1}}$ & $E_{\mathrm{T1}}$ & 
$\Delta U_{\mathrm{GEM2}}$ & $E_{\mathrm{T2}}$ &
$\Delta U_{\mathrm{GEM3}}$ & $E_{\mathrm{T3}}$ &
$\Delta U_{\mathrm{GEM4}}$ & $E_{\mathrm{Ind}}$ \\ 
(V) & (kV/cm)& (V) & (kV/cm)& (V) & (kV/cm)& (V) & (kV/cm) \\
\hline
270 & 4 & 230 & 4  & 288 & 0.1 & 359 & 4 \\
\end{tabular}
\end{center}
\end{table}

\subsubsection{Gain and IBF uniformity}
\begin{figure}[t]
\centering
\includegraphics[width=0.485\textwidth]{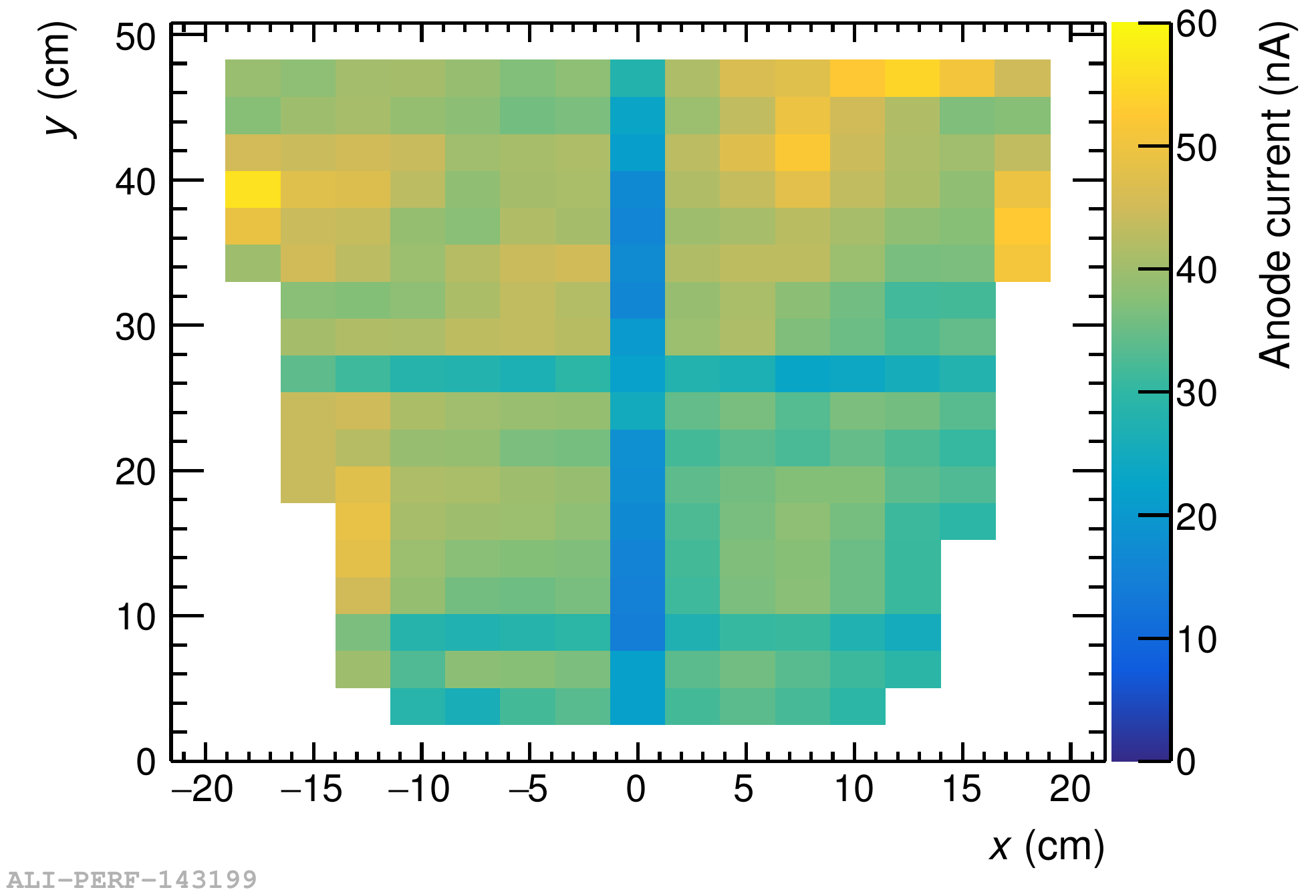}
\includegraphics[width=0.485\textwidth]{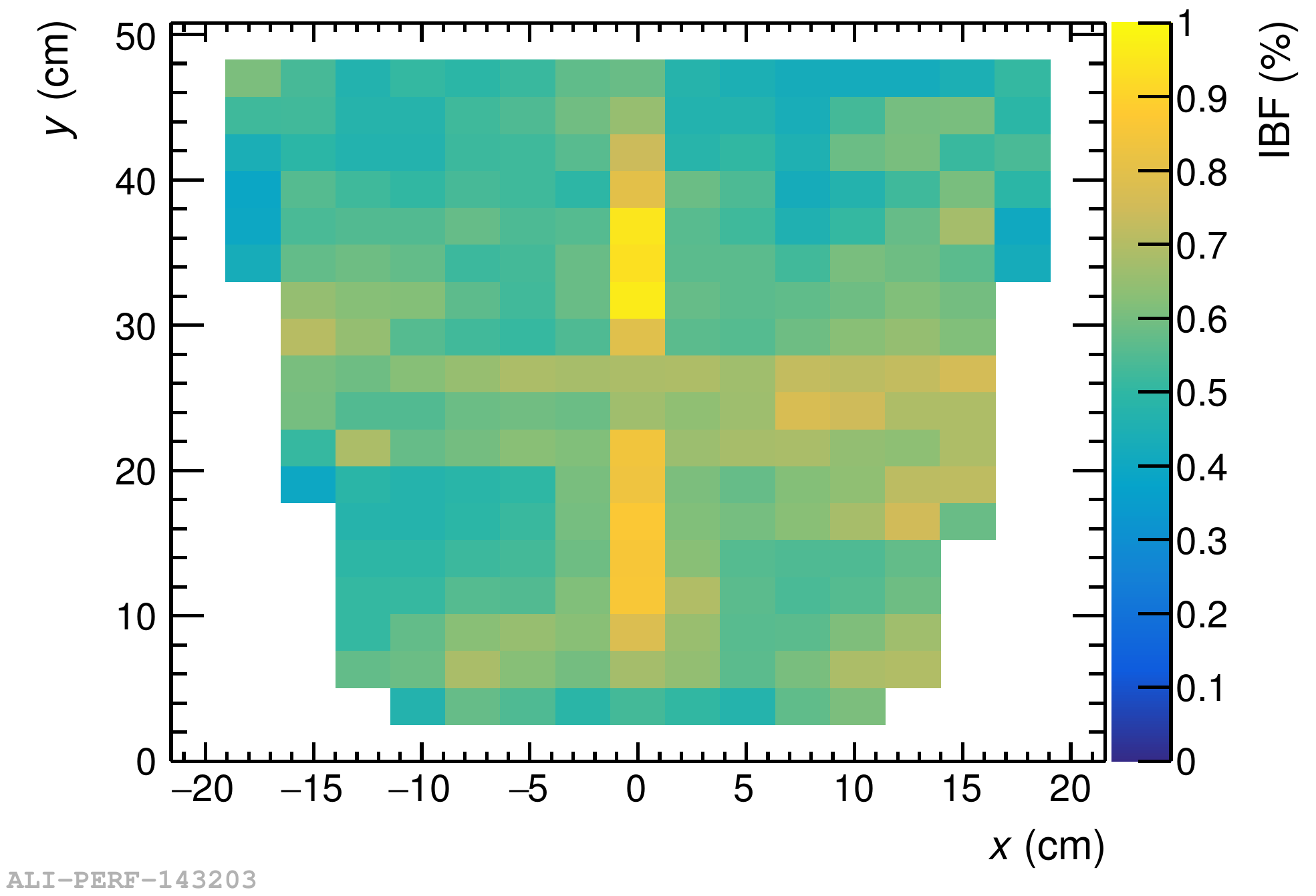}
\caption{Results of the gain (\textit{left}) and IBF (\textit{right}) uniformity tests of the IROC. The cross-shaped structure with low gain corresponds to the supporting grid in the GEM frame. 
%At the edge of the readout chamber, the measured current and hence the gain is reduced due to acceptance effects.
}
\label{fig:IROCresults}
\end{figure}

In order to quantify the uniformity of gain and IBF, all pads are connected together to an amperemeter and additionally the cathode current is monitored. 
The gain uniformity should be better than 20\% RMS and the nominal IBF = 0.7\% with the corresponding variation better than \SI{20}{\percent} RMS.
The detector is irradiated with a collimated $^{55}$Fe source or a moderately powered X-ray generator, whilst operated at the nominal gain of 2000.
Until the GEMs are fully charged up, i.e.\ until the pad current has reached its asymptotical value, the full chamber is irradiated.
Then, the position of the radiation source is varied in steps of $\sim$\SI{2.5}{\centi\meter} and the pad and cathode currents are measured, until the full active area is scanned.
Exemplarily depicted in Fig.~\ref{fig:IROCresults} are the results for the pre-production IROC, which fully conform with the requirements.

\subsubsection{Full X-ray irradiation}
The last qualification step is a stress-test at the maximal current density expected at nominal LHC operation of 10\,nA/cm$^{2}$.
The detector is exposed for at least six hours to the corresponding X-ray flux at the nominal gain, while monitoring the pad current for discharges. After the test, the leakage currents of all GEMs are measured at 250\,V and compared to the situation before the test.

\section{Summary}
During the LS2 of the LHC until 2020 the ALICE TPC will be upgraded with GEM-based readout chambers to conform with the LHC operation scenario.
Two GEM-based prototypes of an Inner and an Outer Readout Chamber with the final design of all relevant elements were built and commissioned to verify and exercise the employed technologies and procedures. The successful conclusion of the commissioning marked the beginning of the final production phase of the upgrade.
All readout chambers are scheduled to be assembled until 2019, when the LS2 is supposed to start. The installation and commissioning of the readout chambers will last until the end of 2020.

\section*{Acknowledgments} 
The speaker acknowledges support by the DFG Cluster of Excellence 'Origin and Structure of the Universe' and the Collaborative Research Center 'Neutrinos and Dark Matter in Astro- and Particle Physics' (SFB 1258).


\begin{thebibliography}{99}
\bibitem{ALICETPC} J. Alme \textit{et al.}, (ALICE TPC Collaboration), \textit{The ALICE TPC, a large 3-dimensional tracking device with fast read-out for ultra-high multiplicity events}, NIM A 676 (2010) 316.
\bibitem{TPC1} B. Abelev \textit{et al.} (ALICE Collaboration), \textit{Technical Design Report for the Upgrade of the ALICE Time Projection Chamber}, CERN-LHCC-2013- 020, 2013.
\bibitem{TPC2} J. Adam \textit{et al.} (ALICE Collaboration), \textit{Addendum to the Technical Design Report for the Upgrade of the ALICE Time Projection Chamber}, CERN- LHCC-2015-002, 2015.
\bibitem{GEM} F. Sauli, \textit{GEM: a new concept for electron amplification in gas detectors}, NIM A 386 (1997) 531.
\bibitem{single-mask} M. Villa \textit{et al.}, \textit{Progress on large area GEMs}, NIM A 628 (2011) 182.
\bibitem{RD1} B. Ketzer, \textit{A Time Projection Chamber for High-Rate Experiments: Towards an Upgrade of the ALICE TPC}, NIM A 732 (2013) 237.
\bibitem{RD2} P. Gasik, \textit{Development of GEM-based Read-Out Chambers for the upgrade of the ALICE TPC}, 2014 JINST 9 C04035.
\bibitem{RD3} C. Lippmann, \textit{A continuous read-out TPC for the ALICE upgrade}, NIM A 824 (2016) 543.
\bibitem{RD4} P. Gasik, \textit{Building a large-area GEM-based readout chamber for the upgrade of the ALICE TPC}, NIM A 845 (2017) 222.
\bibitem{QA2} M. Ball \textit{et al.}, \textit{Quality assurance of GEM foils for the upgrade of the ALICE TPC}, 2017 JINST 12 C01081.
\bibitem{QA} J. Brucken, \textit{The GEM QA protocol of the ALICE TPC upgrade project}, these proceedings.

\end{thebibliography}
\end{document}